\documentclass[floatfix,twocolumn]{aastex631}
\usepackage{amsmath}
\usepackage{bm}
\usepackage[normalem]{ulem}

\received{April 12, 2023}
\revised{June 23, 2023}
\accepted{June 29, 2023}

\submitjournal{ApJ}

\shorttitle{A disk-wind model of 3C 273}
\shortauthors{Long et al.}

\graphicspath{{./}{figures/}}

\begin{document}

\title{\vspace{-8mm}Confronting a Thin Disk-Wind Launching Mechanism of Broad-Line Emission in AGN with GRAVITY Observations of Quasar 3C 273}

\author{Kirk Long}
\affiliation{University of Colorado Boulder}
\author{Jason Dexter}
\affiliation{University of Colorado Boulder}

\author{Yixian Cao}
\affiliation{Max Planck Institute for Extraterrestrial Physics}
\author{Ric Davies}
\affiliation{Max Planck Institute for Extraterrestrial Physics}
\author{Frank Eisenhauer}
\affiliation{Max Planck Institute for Extraterrestrial Physics}
\author{Dieter Lutz}
\affiliation{Max Planck Institute for Extraterrestrial Physics}
\author{Daryl Santos}
\affiliation{Max Planck Institute for Extraterrestrial Physics}
\author{Jinyi Shangguan}
\affiliation{Max Planck Institute for Extraterrestrial Physics}
\author{Taro Shimizu}
\affiliation{Max Planck Institute for Extraterrestrial Physics}
\author{Eckhard Sturm}
\affiliation{Max Planck Institute for Extraterrestrial Physics}

\begin{abstract}
Quasars show a remarkable degree of atomic emission line-broadening, an observational feature which, in conjunction with a radial distance estimate for this emission from the nucleus is often used to infer the mass of the central supermassive black hole. The radius estimate depends on the structure and kinematics of this so-called Broad-Line Region (BLR), which is often modeled as a set of discrete emitting clouds. Here, we test an alternative kinematic disk-wind model of optically thick line emission originating from a geometrically thin accretion disk under Keplerian rotation around a supermassive black hole. We use this model to calculate broad emission line profiles and interferometric phases to compare to GRAVITY data and previously published cloud modelling results. While we show that such a model can provide a statistically satisfactory fit to GRAVITY data for quasar 3C 273, we disfavor it as it requires 3C 273 be observed at high inclination, which observations of the radio jet orientation do not support.

\end{abstract}
\section{Background}
\begin{figure}[ht!]
    \begin{minipage}{0.5\textwidth}
    \centering
    \includegraphics[width=1.\linewidth,keepaspectratio]{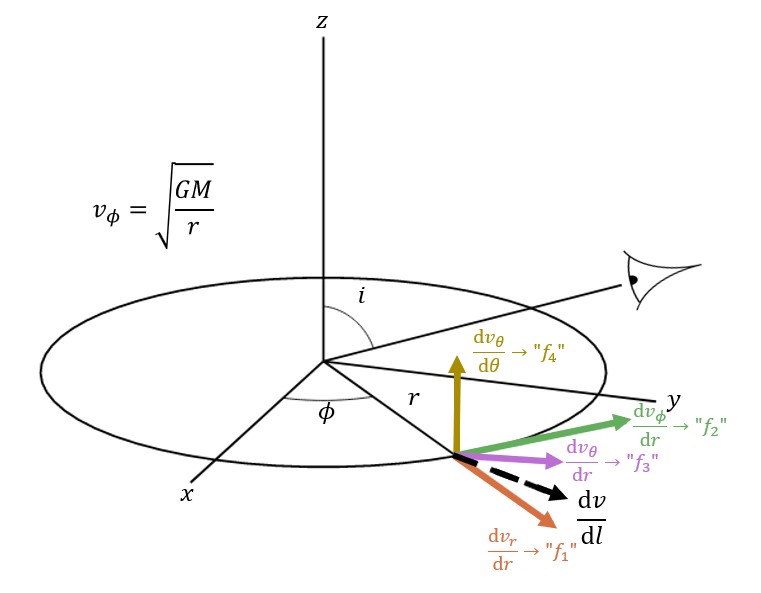}
    \caption{A schematic overview of the model considered in this work. Shown here is one annulus within a Keplerian geometrically thin disk, with the arrows representing various possible velocity gradients that our model incorporates, which we label $f_{1,2,3,4}$. Note that the velocity gradients drawn are just representative possibilities for the many different orientations a wind could take.}
    \label{fig:schematic}
    \end{minipage}
\end{figure}

Quasars are host to sets of emission lines that are single-peaked and have widths on the order of thousands of kilometers per second, scales that are assumed to be set by the gravity of the central supermassive black hole \citep{Peterson2006}. The region where these broad lines originate\textemdash aptly named the ``broad-line region" (BLR)\textemdash is assumed to be at distances of $\sim 10^3-10^4 r_s$ (with the Schwarzschild radius $r_s = \frac{2GM_{\mathrm{BH}}}{c^2}$) from the central black hole, often inferred by measuring time lags between changes in the continuum and the BLR line profiles (see \cite{RM3c27319} for a recent study of this phenomenon in quasar 3C 273). The broad single peak in the line profile is usually explained via a cloud model as in \cite{GRAVITY19} (hereafter G18)\textemdash but \cite{CM96} (hereafter CM96) showed that an alternative ``disk-wind" model could match this morphology in the line profile. If the ratio of outer to inner radius is sufficiently small one would naively expect to observe a double-peak in the line profile, which is often not observed \citep{Jackson1991}. The disk-wind model initially proposed by CM96 avoids this problem by adding in the effects of high velocity gradients present within a Keplerian optically thick geometrically thin disk, with the shears altering the escape probability of photons emitted at different locations in the disk via Sobolev theory \citep{Sobolev}. The high central luminosity of quasars is usually assumed to originate from accretion onto their central supermassive black holes \citep{Rees84,LB69,SS73}, and thus it is tempting to assume this simple disk-like geometry extends to the BLR, and there is significant observational evidence for the presence of winds in quasars \citep{BottorffRM97,Elvis2000QStructure,BALR93}, which could be launched by the large velocity gradients in such a model. Figure \ref{fig:schematic} illustrates the general geometry of the Sobolev disk-wind type model considered in this work.

If we can measure a characteristic size $R_{\mathrm{BLR}}$ and velocity scale $\Delta V$ for orbiting gas in the BLR we can infer the mass of the central supermassive black hole via:

\begin{equation}\label{RMeqn}
    M_{\mathrm{BH}} = f \frac{R_\mathrm{BLR}(\Delta V)^2}{G}
\end{equation}

Here $G$ is the gravitational constant and $f$ is the ``virial factor"\textemdash which originates from assuming that the BLR is virialized and whose value is model dependent on both the geometry and kinematics of the BLR. The notation here matches what is given in \cite{Waters16} (hereafter Waters16). Reverberation mapping techniques allow us to measure a characteristic time delay $t$ between changes in the continuum of the source and changes in the line profile, giving a characteristic size for the BLR of $R_{\mathrm{BLR}}\approx ct$ assuming the change is propagated at the speed of light $c$ \citep{Peterson2006}. GRAVITY spatially resolves the BLR in 3C 273, which provides another method for measuring $R_{\mathrm{BLR}}$. There are significant model and measurement dependent uncertainties in both $R_{\mathrm{BLR}}$ and $\Delta V$, and thus it is of critical importance to constrain what physical models best fit the BLR.

Clear evidence for ordered BLR rotation around 3C 273 (see Figure 1 of G18 or Figure \ref{fig:centroids}) is consistent with both the cloud and the thin disk model. While the cloud model is fit satisfactorily to the data in G18, it has not been considered if a thin disk-wind launching model can also explain the data. \textbf{If the clouds in the model G18 fit to the GRAVITY data are real, how they could survive sufficient time and create sufficient smoothness in the line profiles \citep{Mathews1985,Dietrich1999} is not well understood. A disk-wind model is thus easier to reconcile physically, and even if the clouds are not actual physical objects (and instead just a numerical convenience for fitting and describing the BLR kinematics) it is important to test whether other models can also explain the data.} 

There are many possible disk-wind model morphologies to choose from, but here we consider a simple, two-dimensional optically thick but geometrically thin disk-wind model which is essentially a combination of previous work first explored in CM96 and in a follow up paper published a year later (\cite{MC97}, hereafter MC97) with minor modifications to the velocity gradients and geometries originally considered in those works, a general schematic of which is shown in Figure \ref{fig:schematic}. This model considers a fixed optical depth $\tau$ in the $\tau \gg 1$ limit\textemdash greatly simplifying the equations of radiative transfer\textemdash where an extended thin accretion disk with various hydrodynamical shears drives subtle differences in the escape probability for line photons at different locations in the disk. We include four such shears in our model\textemdash further discussed below\textemdash and allow their strengths to be artificially varied in fitting to determine what possible wind orientations might exist. 

We note that this is not the only possible disk-wind model that could explain this data, as other authors have explored variations on this idea in a variety of capacities: \cite{ChajetHall2013} presented a version of the CM96/MC97 model where they adoped a hydromagnetic prescription for the wind\textemdash as opposed to purely hydrodynamical \textemdash following on the work of \cite{Emmering1992}; \cite{Flohic2012} corrected a minor error in MC97 and added in the effects of relativity, sampling a wide parameter space to produce line profiles that could be compared to observations; and many groups have included variants on the calculation presented in MC97 where the wind has a substantial vertical extent \citep{Waters16,BaskinLaor2018,NaddafCzerny2022}, with some applying Monte-Carlo methods to better model the radiative transfer than we do here \citep{Mathews2020}. We choose the model we do for its flexibility, simplicity, and computational ease of fitting as a first test to GRAVITY data. 
\newpage
\section{Methods}
As in CM96 and MC97, we utilize the Sobolev \citep{Sobolev} approximation in modelling the line emission, as the macroscopic velocities of gas within the disk are much larger than the intrinsic line width. The Sobolev optical depth is proportional to the inverse of the line of sight velocity gradient, which in this work is modelled as various possible shears within the thin disk geometry. In the high optical depth limit within the Sobolev regime the equation of radiative transfer is essentially dominated by the source function multiplied by the probability that a photon will escape the disk, which in the optically thick limit is directly proportional to the line of sight velocity gradient. Thus, any anisotropic terms present in the velocity gradients of the hydrodynamic thin disk can alter the morphology of the observed line profile, leading to a variety of interesting possible shapes outside of the expected double peak.

In the Sobolev approximation we can calculate the frequency dependent line profile from a steady continuum source in a thin-disk $L_\nu$ as:
\begin{equation} \label{eq0}
    L_{\nu} =\sin i \int r\mathrm{d}r\int_0^{2\pi}\mathrm{d}\phi k(r)\beta (r,\phi,i)S(r)\delta[\nu-\Tilde{\nu}(\phi,r)]
\end{equation}
Here $i$ is the viewing inclination angle of the system, with $r$ and $\phi$ disk coordinates as shown in Figure \ref{fig:schematic}. $k$ is the integrated line opacity (proportional to the density of the emitting gas), $\beta$ the photon escape probability, and $S$ the source function, which here is modelled as a power-law function of $r$ alone\textemdash $S(r) \propto r^{-\alpha}$ with $\alpha = 1$ in keeping with CM96. This relationship is essentially the familiar equation for the formation of a spectral line, but the usual line function is replaced with the Dirac-delta function in keeping with the Sobolev approximation. 

This relationship was first shown by \cite{RH83}, and here this is the same as CM96's equation 8, where (as shown in Waters16) $\Tilde{\nu} = \nu_0\left(1+\frac{v_l}{c}\right)$ is the Doppler-shifted resonant frequency of the line as seen by the observer, i.e. $v_l = \sin i(v_r\cos\phi - v_\phi\sin\phi) = -\sin iv_\phi\sin\phi$ ($v_r = 0$ and the $l$ subscript here denotes that this quantity is along the observer's line of sight). Note that this is only the classical Doppler shift from circular motion, and does not include the effects of relativity or transverse motion. Both effects are radially dependent and act to make the line profile more skewed, with the relativistic effect being the more important of the two with a scaling on the order of $1.5(r_s/r)\times 10^5$ km/s (Waters16). Fortunately, the characteristic size of the BLR is of order $10^3r_s$ and thus we do not model this effect as it is not significant over the width of the line profile. \cite{Flohic2012} have explored the effects of relativity on a similar disk-wind model and have computed line profiles for a wide and robust set of parameters, which verify that for the parameter space surveyed in this work relativity is not a primary factor in the shape of the line profiles.

If we also consider continuum variability, there will be an associated time lag at each disk location in addition to the Doppler shift. Adding this in gives us the so-called ``transfer function" (as shown in Waters16 and originally introduced by \cite{BlandfordMcKee1982}):

\begin{equation} \label{eq01}
    \Psi(t,\nu) = \iint r k(r)\beta (r,\phi,i)S(r)\delta[\nu-\Tilde{\nu}(\phi,r)]\delta[t-\Tilde{t}]\mathrm{d}r \mathrm{d}\phi
\end{equation}

For a thin disk with Keplerian rotation this resonance condition for the time delay $\Tilde{t}$ is given by $\Tilde{t} = \frac{r}{c}(1-\cos\phi\sin i)$. 
The de-projected line profile is then just $\Psi(\nu)=\int_0^\infty\Psi(t,\nu)\mathrm{d}t$ and the so-called response function is $\Psi(t)=\int_{-\infty}^\infty\Psi(t,\nu)\mathrm{d}\nu$, which we illustrate with echo images of our best fit in the results section. 

Using the standard Sobolev approach we express the escape probability as $\beta = \frac{1-e^{-\tau}}{\tau}$\textemdash with $\tau$ here representing the line optical depth, given by $\tau =\left(\frac{k(r)c}{\nu_0}\right)\lvert\frac{\mathrm{d}v_l}{\mathrm{d}l}\rvert^{-1}$. Here we consider optically thick line emission to simplify the equations of radiative transfer (i.e. $\tau \gg 1$), and in this limit the escape probability $\beta$ reduces to simply $\beta \approx \frac{1}{\tau}$. In the Sobolev approximation $\tau$ is inversely proportional to the line of sight velocity gradient, giving $\beta \approx \frac{1}{\tau} = \left[\frac{\nu_0}{k(r)c}\right]\lvert\frac{\mathrm{d}v_l}{\mathrm{d}l}\rvert$. Thus we can express the quantity $k(r)\beta(r) = \frac{\nu_0}{c}\lvert\frac{\mathrm{d}v_l}{\mathrm{d}l}\rvert$. The core assumptions of this approximation are that it is hard for line photons in general to escape the disk as the line is optically thick, with the probability for escape being shaped by the velocity gradients (shears) present within the disk. This allows for more complicated line morphologies than one might naively expect.

As discussed above we choose a simple power-law dependence for the source function to match previous work, but note that as shown in Waters16 one can think of the source function as going like $S(r) = A(r)F_X^{\eta(r)}$, where $F_X$ is the flux from the continuum and the power law index $\eta(r) = \frac{\partial \ln S_l}{\partial \ln F_X}$ (where $S_l$ is the source function along the line of sight, for further details see \cite{Krolik91}). Typical $\eta$ values from photoionization modeling are between 0 and 2, with $A(r) = A_0r^\gamma$ setting the overall response amplitude in the line. Equation \ref{eq0} shows that the emissivity in the line $j \propto S(r)$, thus the responsivity of the line is $\frac{\partial j}{\partial F_X} \propto A(r)\eta(r)F_X^{\eta(r) - 1}$. In this formalism the values chosen for $\eta$ and $\gamma$ then set the overall radial scaling of the source function as well as how the line emission responds to the continuum, with higher values of $\gamma$ enhancing the response at larger radii. Note that there are then many possible models of how the BLR responds to illuminating continuum flux on the disk for a given choice for $S(r)$. Assuming illumination from a central source onto a thin disk BLR gives $F_X\propto r^{-3}$, thus for the constraints on $\eta$ we can constrain $\gamma$ to be between $-1$ (for $\eta = 0$) and $5$ (for $\eta = 2$) to yield the source function $S(r) \propto r^{-1}$ we use in this work.

Applying this prescription for the source function as well as the simplification for $k(r)\beta(r)$ then reduces equation \ref{eq01} to:

\begin{equation} \label{eq02}
    \Psi(\nu,t) = \Psi_0\frac{\nu_0}{c}\int_{r_{\mathrm{min}}}^{r_{\mathrm{max}}}\int_0^{2\pi}\lvert\frac{\mathrm{d}v_l}{\mathrm{d}l}\rvert\delta[\nu-\Tilde{\nu}]\mathrm{d}r\mathrm{d}\phi\delta[t-\Tilde{t}]
\end{equation}

Where $\Psi_0$ is a normalization constant, whose value one would need to specify to compare to physical flux values but here we leave as an arbitrary constant as we are only attempting to match the shape of the line profile.

We will evaluate this integral numerically with the following general approach:
\begin{enumerate}
    \item First, we make a 2D disk in log polar coordinates, where each cell has coordinates ($r$, $\phi$) that correspond to an associated resonant Doppler shift $\Tilde{\nu}$ and time delay $\Tilde{t}$. The general geometry of the model is shown in Figure \ref{fig:schematic}.
    \item We then calculate the intensity escaping towards the observer at each location within the disk as given by Equation \ref{eq02}.
    \item Finally, we integrate to get the total line luminosity ($\Psi{\nu} = \int_0^\infty \Psi(\nu,t)\mathrm{d}t$), binning the disk according to $\Tilde{\nu}$ and summing over all time delays to get the line profile and vice versa ($\Psi{t} = \int_{-\infty}^\infty \Psi(\nu,t)\mathrm{d}\nu$) to obtain the response function. This ensures each region of the disk only contributes to the total intensity integral at its corresponding resonant Doppler frequency as is required by the Dirac-delta function in Equations \ref{eq0}-\ref{eq02}. 
\end{enumerate}

We must now evaluate $\lvert\frac{\mathrm{d}v_l}{\mathrm{d}l}\rvert$. The line of sight velocity gradient can be found using the rate of strain tensor\newline $\hat{n}\cdot\boldsymbol{\Lambda}\cdot\hat{n}$ (often denoted as $Q = \sum\limits_{i,j}\frac{1}{2}\left(\frac{\partial v_i}{\partial r_j} + \frac{\partial v_j}{\partial r_i}\right)$)\textemdash  shears in the disk create the velocity gradients we seek to recover, which we will now derive. The $\hat{n}$ for this geometry at an observer of $\phi = 0$ and inclination $i$ is given by Waters16 (see there and/or CM96, \cite{Flohic2012} for further discussion and derivation) as:

\begin{equation}\label{nhat}
\begin{split}
    \hat{n} = & (\sin\theta\cos\theta\sin i + \cos\theta\cos i)\hat{r} \\
    & + (\cos\theta\cos\phi\sin i - \sin\theta\cos i)\hat{\theta} \\
    & - (\sin\phi\sin i)\hat{\phi}
\end{split}
\end{equation}

Using this $\hat{n}$ and the rate of strain tensor terms in spherical coordinates from \cite{Batchelor68} we can write the line of sight velocity gradient as:

\begin{equation} \label{eq1}
\begin{split}
\hat{n}\cdot\boldsymbol{\Lambda}\cdot\hat{n} = & \sin^2i\bigg[\frac{\partial v_r}{\partial r}\cos^2\phi - \left(\frac{\partial v_\phi}{\partial r} - \frac{v_\phi}{r}\right)\sin\phi\cos\phi\\ 
& + \frac{v_r}{r}\sin^2\phi \bigg] \\
& -\sin i\cos i\bigg[\left(\frac{1}{r}\frac{\partial v_r}{\partial \theta} + \frac{\partial v_\theta}{\partial r} - \frac{v_\theta}{r}\right) \cos\phi\\
& - \frac{1}{r}\frac{\partial v_\phi}{\partial \theta}\sin\phi\bigg] \\
& +\cos^2i\left[\frac{1}{r}\frac{\partial v_\theta}{\partial \theta} + \frac{v_r}{r}\right]
\end{split}
\end{equation}

Where in arriving at the form above we have assumed all of the $\frac{\partial}{\partial \phi}$ operator terms are 0 (the underlying disk is Keplerian) and the disk is in the equatorial plane ($\theta = \frac{\pi}{2}$) which allows us to significantly simplify the result. A more thorough derivation of equation \ref{eq1} is given in the Appendix A. We keep the $\theta$ terms here to generalize the wind in both the ``vertical" and radial directions. This is the same result as given in CM96/MC97 but with the $\phi$ convention for the observer given in Waters16, which differs from CM96/MC97 by $-\frac{\pi}{2}$. For the rest of this work we will use the $\phi$ convention specified in Waters16 (equation \ref{eq1}).

In CM96/MC97 they assume that $v_r \approx 0$ in the thin disk, but that there is an acceleration related to the escape velocity, ie $\frac{\partial v_r}{\partial r} \approx 3\sqrt{2}\frac{v_\phi}{r}$, where $v_\phi = \sqrt{\frac{GM}{r}}$ is the Keplerian $v_\phi$, which gives us $\frac{\partial v_\phi}{\partial r} =  \frac{-v_\phi}{2r}$. These accelerations are important, as in Sobolev theory the velocity gradients essentially give us the escape probability of photons resonating in the 
thick medium of the disk, and it is these escaping photons that we image \citep{Sobolev,RH83}.

But what are the $\theta$ terms? Following in the footsteps of CM96/MC97 it makes sense to assume that on average $v_\theta \approx 0$ for the same reason $v_r \approx 0$, but similarly we will assume a particle may be lifted by the wind and accelerated to the local escape velocity (but now in the $\hat{\theta}$ direction) such that $\frac{\partial v_\theta}{\partial \theta} \approx \frac{v_{esc}}{\left(H/R\right)}$ (where $H/R\ll 1$ is the scale height of the disk) and $\frac{\partial v_\theta}{\partial r} \approx \frac{\partial v_r}{\partial r}$. Since $v_\phi(r)$ is a function of $r$ alone $\frac{\partial v_\phi}{\partial \theta} = 0$, and we also set $\frac{\partial v_r}{\partial \theta} = 0$ in keeping with the idea of a thin disk.

These approximations reduce equation \ref{eq1} to:

\begin{equation} \label{eq3}
\begin{split}
\hat{n}\cdot\boldsymbol{\Lambda}\cdot\hat{n} = & 3\frac{v_\phi}{r} \sin^2i\cos\phi\left[\sqrt{2}\cos\phi + \frac{\sin\phi}{2}\right]\\
& -\sin i \cos i \left[3\sqrt{2}\frac{v_\phi}{r}\cos\phi\right] \\
& +\cos^2i\left[\frac{1}{r}\frac{v_{esc}}{(H/R)}\right]
\end{split}
\end{equation}

Rescaling $v_\phi$ into units of $r_s$ gives us $v_\phi = \sqrt{\frac{1}{2r'}}$ and $v_{esc} = \sqrt{\frac{1}{r'}}$ (where $r' = r/r_s$, so $r = r'r_s$), which, after simplifying, gives us:

\begin{equation} \label{eq4}
\begin{split}
\hat{n}\cdot\boldsymbol{\Lambda}\cdot\hat{n} = & \frac{1}{r_s}\sqrt{\frac{1}{2r'^3}} \bigg [3 \sin^2i\cos\phi\left(\sqrt{2}\cos\phi + \frac{\sin\phi}{2}\right)\\
& -3\sqrt{2} \sin i \cos i \cos\phi +\frac{\sqrt{2}}{(H/R)}\cos^2i \bigg ]
\end{split}
\end{equation}

There are four different possible $\phi$ dependencies (within the angle brackets) as a result of these shears within the disk, which we qualitatively describe below:
\begin{enumerate}
    \item The first term ($f_1$ in Figure \ref{fig:schematic}, $\propto \cos^2\phi$) largely describes radial shear from the wind, where the angular dependence allows photons with small Doppler shifts to escape more easily to the observer from regions with large radial shears at the near and far sides of the disk ($\phi \approx 0$ or $\pi$, where the line of sight projected velocities are small).\vspace{-1mm}
    \item The second term ($f_2$ in Figure \ref{fig:schematic}, $\propto \cos\phi\sin\phi$) describes gradients caused by Keplerian shear in the disk, which by itself produces a double-peaked line profile (with peaks corresponding to the blue and red sides of the disk), replicating the ``M profile" first shown in \cite{RH83} and further discussed in the context of double-peaked line profiles from cataclysmic variables by \cite{HorneMarsh1986}. \vspace{-1mm}
    \item The third term($f_3$ in Figure \ref{fig:schematic}, $\propto \cos\phi$) represents the ``lifting" shear as a function of radius, where again the angular dependence allows photons with small Doppler shifts to escape more easily (although less strongly than in the case of radial shear).\vspace{-1mm}
    \item The final term ($f_4$ in Figure \ref{fig:schematic}) represents the ``lifting" shear as a function of height off of the disk, and it interestingly has no $\phi$ dependence, meaning it represents a form of isotropic emission that by itself produces a doubly peaked line profile, albeit of a different shape than the profile given by just the Keplerian shear. For thin disks $H \ll R$ and thus even a small velocity gradient in the $\theta$ direction will be amplified greatly by this term, so to keep its magnitude similar to the other terms we absorb this dependence into $f_4$ (i.e. $f_4 = \frac{C}{(H/R)}$ where $C$ is the unamplified wind contribution in this direction).\vspace{-1mm}
\end{enumerate}

\textbf{This is the crux of our model, as the emission intensity at each location in the disk is set by $\lvert\frac{\mathrm{d}v_l}{\mathrm{d}l}\rvert$ alone in the optically thick limit, as shown in equation \ref{eq02}.}

\begin{figure*}[!ht]
    \centering
    \includegraphics[width=\textwidth,height=\textheight,keepaspectratio]{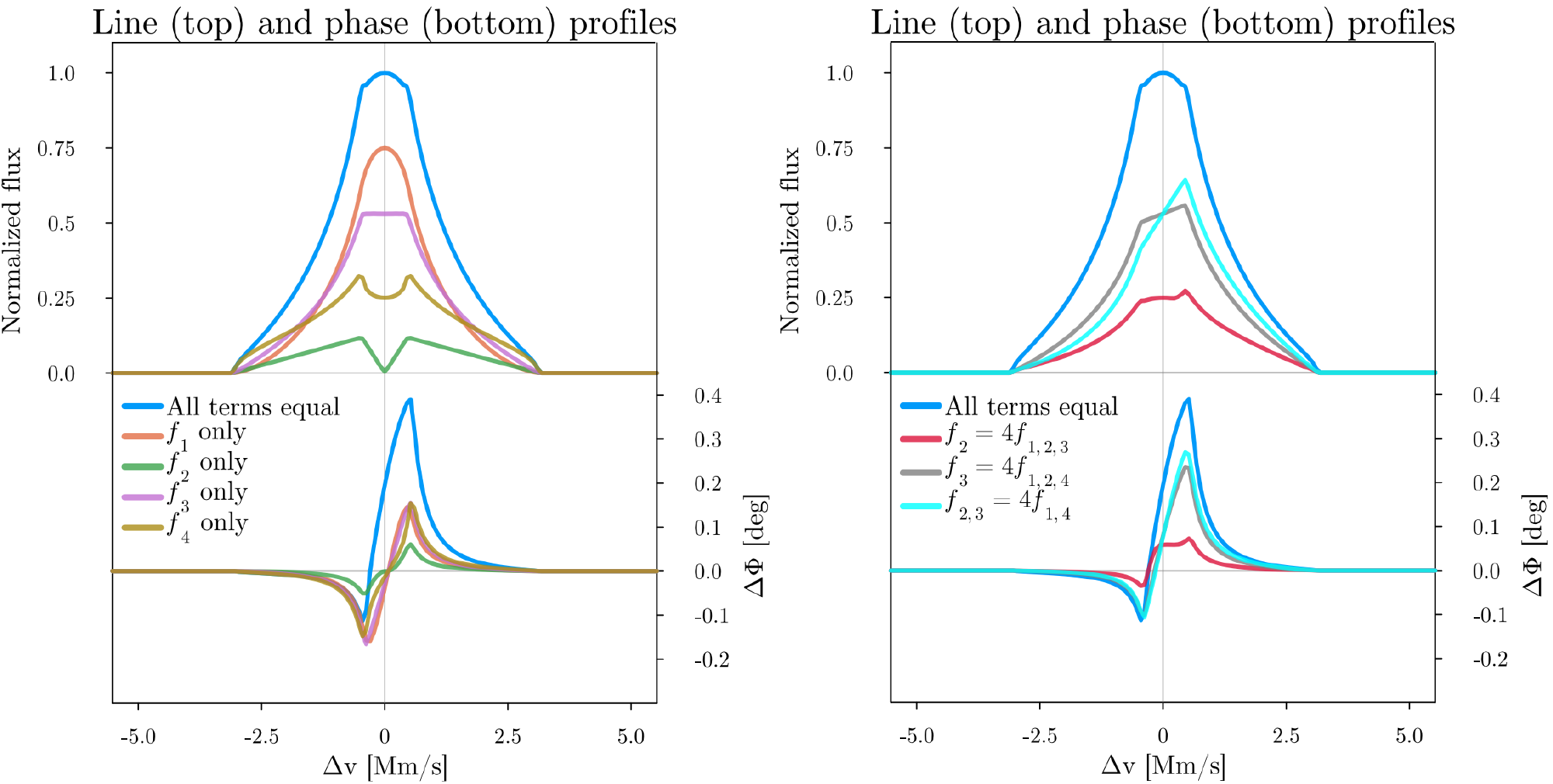}
    \caption{Sample line (top) and phase (bottom) profiles for several combinations of wind terms from our model viewed at an inclination of $45^\circ$, with other parameters chosen to roughly match those fit to in the results section. The left panel shows how each term acts on its own, and we see that terms $f_1$ and $f_3$ produce symmetric, single peaked line profiles, while terms $f_2$ and $f_4$ create double-peaked profiles. In conjunction with other terms $f_2$ and $f_3$ create a red-blue asymmetry about the line center, which is highlighted in the right panel. The legend indicates the strengths of the terms with respect to each other, as only the ratio of the strengths matters in our model, with the ``all terms equal" line representing output from a sample disk with $f_1 = f_2 = f_3 = f_4 = 1$.}
    \label{fig:fig1}
\end{figure*}

To better explore the parameter space we want to examine how each of these terms affect the line profile shape. Equation \ref{eq02} is directly proportional to the line of sight velocity gradient\textemdash a result of our assumption that $\tau \gg 1$ in the Sobolev approximation\textemdash and this makes it the critical component that shapes the line and response profiles. Discarding the normalization and primes, we can explore how this line of sight velocity gradient changes as a function of $r$, $i$, and $\phi$ alone:

\begin{equation} \label{eq6}
\begin{split}
\hat{n}\cdot\boldsymbol{\Lambda}\cdot\hat{n} \approx \frac{\mathrm{d}v_l}{\mathrm{d}l} \approx & \sqrt{\frac{1}{2r^3}} \bigg (3\sin^2i\cos\phi\left[\sqrt{2}{f_1}\cos\phi + {f_2}\frac{\sin\phi}{2}\right]\\
& - 3 {f_3} \sin i \cos i \cos\phi + \sqrt{2}{f_4} \cos^2i\bigg)
\end{split}
\end{equation}

Neglecting physical constants, this completes the mathematical description of our model, where ${f_{1,2,3,4}}$ are artificial constants that allow the fit to be more flexible, varying the strength of the different wind launching terms described above. In the high inclination limit only the $\sin^2i$ term is important and our result reduces to the form considered in CM96, and overall our result is similar to that presented in MC97. At low inclinations the $\cos^2i$ term is very important, and the addition of both terms at moderate inclinations make a significant difference when compared with the results shown in CM96 as demonstrated originally in MC97.

Figure \ref{fig:fig1} explores the general behavior of these terms and how they affect the shape of the line and phase profiles. Qualitatively, the terms have the following effects on the line profile:
\begin{enumerate}
    \item ${f_1}$ creates a symmetric bell-curve like shape about the line center, as it is proportional to $\cos^2\phi$.\vspace{-2mm}
    \item ${f_2}$ in isolation creates an ``M" shaped line profile with two peaks, with the line center as a minimum. This replicates the result first shown in RH83's figure 2. In conjunction with other terms it acts to  depress the region just to the left of the center of the line profile, and to raise the region just to the right of the center as it is proportional to $\cos\phi\sin\phi$, providing a red-blue asymmetry to the line profile. \vspace{-1mm}
    \item ${f_3}$ in isolation creates a flat-topped line profile, with the plateau centered on the line center, but in conjunction with other terms it acts to widen the line profile and alter the shape of the peak, as it is proportional to $\cos\phi$ and in our definition the observer is at $\phi=0$, meaning there is no left/right asymmetry. \vspace{-1mm}
    \item ${f_4}$ creates a symmetric ``twin-horned" type feature, as it is an isotropic emission feature and thus only depends on the Doppler shift, i.e. the frequency shift is proportional only to $\sqrt{\frac{1}{2r}}\sin i\cos\phi$. This is analogous to the dash-dot profile shown in CM96's figure 2. Its shape is thus entirely dependent on the delta function Doppler resonance\textemdash gas on the bluer/redder sides of the disk resonates in blue/red wavelengths and thus only contributes on the corresponding side of the line profile, while the gas with no Doppler shift close to the front and back sides of the disk keeps the line profile from going to zero at line center. 
\end{enumerate}

Regardless of whether the individual term would produce a single or double peaked line profile, all of the phase profiles display the standard ``S" shape, but with different morphologies. This pattern in the phase profile indicates a red-blue asymmetry in the emission line centroids, as expected for ordered rotation, with the differential phase being a measurement made possible through interferometry with the GRAVITY instrument on the VLTI \citep{GRAVITY17}. As shown in the left panel of figure \ref{fig:fig1}, when all the terms of are of equal strength ${f_1}$ and ${f_4}$ dominate. 

Our code uses this model prescription to generate a two-dimensional ray-traced image of an inclined disk assumed to represent the BLR, employing a polar coordinate grid with logarithmic radial spacing. The intensity at each grid cell in the disk is given by equations \ref{eq02} and \ref{eq6}, and we can calculate the line profile numerically as described above (with the summed intensities at each disk location weighted by their corresponding area element). In fitting we allow the terms to vary independently with priors $0 \leq {f_{1,2,3,4}} \leq 1$ consistent with the disk launching outflows. The differential phase is calculated following the standard BLR photocenter and kinematic modelling prescription in the marginally resolved limit, where we keep only the first order term in the expansion of the complex visibility such that $\Delta \Phi = -2\pi\left(u\cdot\bar{x}\right)\frac{f}{1+f}$  \citep{GRAVITY35int,GRAVITY36int}. Here $u$ are the interferometric baselines, $\bar{x}$ the on-sky emission centroids, and $f$ the normalized line flux such that $\frac{f}{1+f}$ represents the contrast between the line and the continuum. Note that this $\Delta \phi$ does not refer to any physical $\phi$ in the disk and is only the differential phase angle. 

We use data previously published by the GRAVITY collaboration on the quasar 3C 273 to fit our model\textemdash for a detailed description of the observations and reduction techniques used, see G18. 3C 273 is uniquely suited to test the model because the system's distance is close enough that GRAVITY obtains a spectroastrometric differential phase signature across the broad line profile which we can fit for in addition to the shape of just the line profile \citep{GRAVITY17,GRAVITY19}. 3C 273 is also oriented such that we observe the jet, allowing us to constrain what we believe the ``true" inclination of the system to be \citep{jetAngle86}. The fitting is done with flux and phase measurements along each wavelength channel (measurements taken in 40 channels between $\sim$ 2.13 and $\sim$ 2.22 $\mu$m over six baselines at four different epochs), using Markov Chain Monte Carlo (MCMC) methods. To ensure we sample a large region of the parameter space we employ a parallel tempered MCMC method developed by \cite{ptemcee16,emcee13}, using six different logarithmically spaced temperatures each with 24 walkers. The model as we fit to the data is fully described by the following 11 parameters:
\begin{enumerate}
    \item The inclination angle of the system $i$, where $i = 90^\circ$ corresponds to an edge-on viewing angle and $i = 0^\circ$ a ``face-on" viewing angle as shown in \ref{fig:schematic}. Higher values of $i$ lead the first term in equation 6 to be the dominant drivers of the line and phase profiles, with lower values leading the last term to be most significant. 
    \item The mass of the central supermassive black hole, $M_{\textrm{BH}}$. Increasing the mass of the black hole increases the amplitude of the phase profile.
    \item The mean radius of the BLR as weighted by the emissivity $j(r) = k(r)\beta(r)S(r)\approx\frac{\nu_0}{c}\lvert\frac{\mathrm{d}v_l}{\mathrm{d}l}\rvert S(r)$, which for our scaling of $S(r)$ gives $j(r)\propto r^{-5/2}$ and thus:
    $$\bar{r} = \frac{\int\limits_{r_{\textrm{min}}}^{r_{\textrm{max}}} r j(r) \mathrm{d}r}{\int\limits_{r_{\textrm{min}}}^{r_{\textrm{max}}} j(r) \mathrm{d}r} = 3\left(\frac{r_{\textrm{max}}^{-1/2}-r_{\textrm{min}}^{-1/2}}{r_{\textrm{max}}^{-3/2}-r_{\textrm{min}}^{-3/2}}\right)$$ 
    Higher values of $\bar{r}$ lead to the line/phase profiles being ``squeezed" in wavelength space.
    A size scaling factor $r_{\textrm{fac}}$, which in conjunction with $\bar{r}$ gives the minimum and maximum radii of the BLR via $r_{\textrm{min}} = \frac{\bar{r}}{3}\left(\frac{r_{\textrm{fac}}^{-3/2} - 1}{r_{\textrm{fac}}^{-1/2}-1}\right)$ and $r_{\textrm{max}} = r_{\textrm{fac}} r_{\textrm{min}}$. Increasing $r_{\textrm{fac}}$ slightly ``stretches" the line/phase profiles in wavelength space, and also steepens the slope in the S-curve of the phase profile connecting the negative and positive peaks. This also increases the total flux of the line, but this doesn't affect the fit as we are only seeking to match the characteristic line shapes.
    \item The proportional strength of the radial shear wind term, ${f_1}$.
    \item The proportional strength of the Keplerian shear term, ${f_2}$.
    \item The proportional strength of the radial lifting shear term, ${f_3}$.
    \item The proportional strength of the height lifting shear term, ${f_4}$.
    \item The rotation of the model with respect to the orientation of the baselines in the data, $\theta_\mathrm{PA}$, reported in the standard convention in reference to the orientation of the jet (90$^\circ$ offset from the disk).
    \item A parameter $n$ that can vary the normalization of the line profile with respect to the data slightly as the data points may not be exactly at the peak of the line, $n \ge 1$, where $n = 1$ corresponds to scaling the model exactly to the maximum flux measurement in the data.
    \item A parameter $\Delta \lambda_c$ that varies the line center, thus slightly shifting the models left and right in $\lambda$ space. Here we model Pa $\alpha$ line emission, which has a known center near $\lambda_c \approx 2.172 \mu \mathrm{m}$ for 3C 273 at a redshift of $\sim 0.16$. 
\end{enumerate}
\vspace{-5mm}

\begin{deluxetable*}{ccccccccccc}\label{table1}
\tabletypesize{\scriptsize}
\tablewidth{0pt} 
\tablecaption{MCMC fit parameters}
\tablehead{
\colhead{\large$i \ [^\circ]$} & 
\colhead{\large$M_{\textrm{BH}} \ [10^7 M_\odot]$} & 
\colhead{\large$\bar{r} \ [r_s]$} & 
\colhead{\large$r_{\textrm{fac}}$} & 
\colhead{\large${f_1}$} & 
\colhead{\large${f_2}$} &
\colhead{\large${f_3}$} & 
\colhead{\large${f_4}$} & 
\colhead{\large$\theta_{\textrm{PA}} \ [^\circ]$} & 
\colhead{\large$n$} & 
\colhead{\large$\Delta\lambda_c \ [\mu \rm{m}]$}
}
\colnumbers
\startdata
${78^{+4.9}_{-15}}$ & $8.4^{+0.93}_{-1.7}$ & $6400^{+320}_{-1100}$ & $47^{+3.3}_{-7.6}$ & $0.69^{+0.14}_{-0.32}$ & $0.80^{+0.11}_{-0.53}$ & $0.65^{+0.17}_{-0.51}$ & $0.44^{+0.23}_{-0.40}$ & ${240^{+3.9}_{-8.1}}$ & $1.0^{+0.005}_{-0.008}$ & $0.00032^{+0.0}_{-0.0}$\\
\enddata
\tablecomments{Means with 1$\sigma$ percentile confidence intervals (to two significant figures) on each of our 11 fit parameters, each of which is fully described on the previous page (the ordering 1-11 matches the ordering of the table). ${f_3}$ and ${f_4}$ are particularly poorly constrained, a result of the higher inclination preference of the sampler.}
\end{deluxetable*}
\vspace*{-5mm}

\vspace{-1mm}
\newpage
\section{Results}
After fitting the model as described in the previous section, we reached convergence after $\sim$ 10,000 iterations in each walker at each temperature. We consider the fit converged when the maximum autocorrelation time of any parameter is 1\% of the total number of steps taken. We show only the lowest temperature in our results presented below as the upper temperatures are designed to explore the parameter space and ``trickle" down to the lowest temperature for further refinement \citep{ptemcee16}. Most importantly, the fit generally prefers higher inclinations and thus lower black hole mass. Table \ref{table1} presents the mean values in the fit with 1$\sigma$ confidence intervals.

\begin{figure}[ht]
    \begin{minipage}{0.47\textwidth}
    \includegraphics[width=1.\linewidth,keepaspectratio]{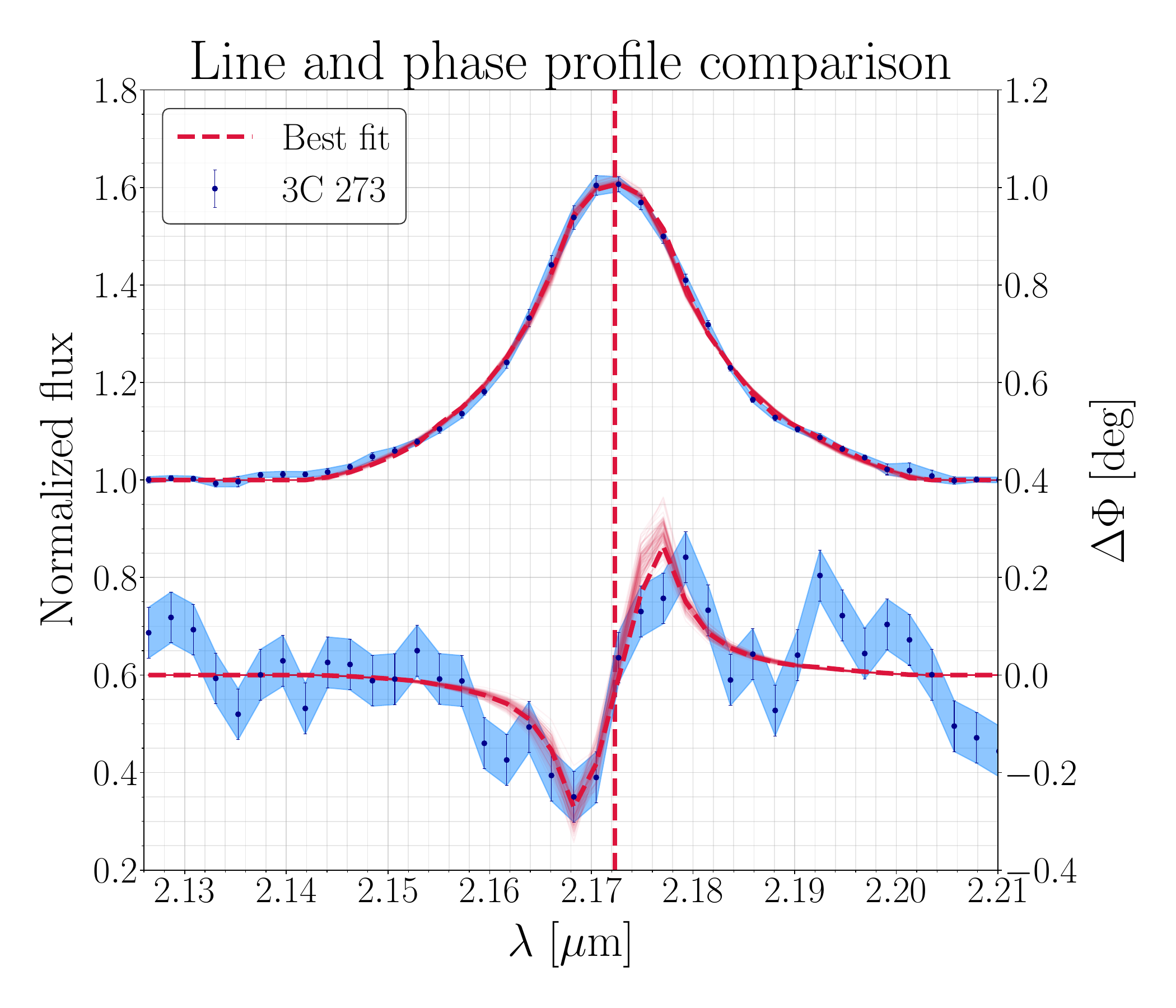}
    \caption{The resulting line (top) and phase (bottom) profiles from our best fit, with the full parameter list contained in table 1. The dashed red line is the best fit from the model (with the vertical dash indicating the line center), while the translucent red lines are 100 random draws from the sampler illustrating the spread of the fit. The phase profile shown here is an average of the phases recovered from the baselines which are significantly misaligned with the jet axis (baselines UT4-UT1, UT4-UT2, and UT4-UT3\textemdash the same as in G18). Figures \ref{fig:fullphase} and \ref{fig:baselines} in our appendix show all of the individual phase profiles and the uv coverage of the observations, respectively.}
    \label{fig:best_fit}
    \end{minipage}
\end{figure}

Figure \ref{fig:best_fit} shows our best fit to the line and phase profiles, with the fainter red lines showing the distribution of the samples represented in table \ref{table1}. The appendix shows a corner plot (Figure \ref{fig:corner}) of all of our parameters with their associated one-dimensional histograms. The fit is good, with a reduced $\chi^2$ value of $\sim 1.36$ for the mean parameters (and $\sim 1.35$ for the best fit). The fit prefers higher inclinations, leading to only the ${f_{1,2}}$ terms being significant in the fit. The uncertainty on the importance of the wind terms with respect to each other is large and thus it is difficult to draw conclusions on the importance of any aspect over another, aside from the larger importance in ${f_{1,2}}$ that is largely driven by the inclination dependence. However this does appear to match observational evidence that there may be a large radial velocity component in any disk-wind outflows \citep{Vestergaard2000}. 

\begin{figure*}
    \centering
    \includegraphics[width=\textwidth,height=\textheight,keepaspectratio]{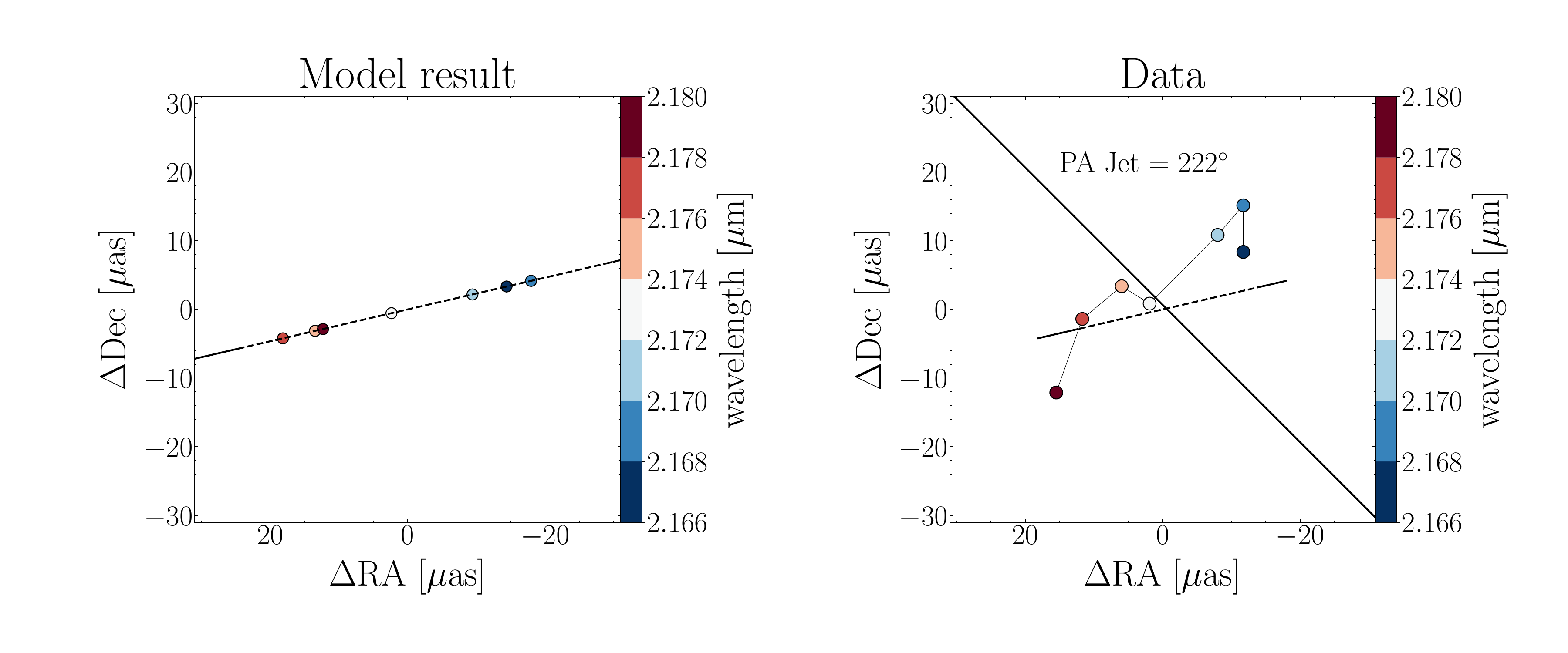}
    \caption{Centroids from our best-fit model sampled at the same wavelengths in the data are shown in the left panel, with a best fit line representing the centroid track. The right panel shows the data centroids with the solid black line representing the PA of the jet as shown in G18 and the dashed line corresponding to the best fit centroid track from our model. Note that the $\bar{r}\approx 17\mu$as in our fit essentially matches the extent of the data centroids shown as our model is a disk governed by Keplerian rotation, while the result in G18 corresponds to a significantly larger mean size for the BLR of $\sim 50 \mu$as, illustrating that the emission centroids showing ordered rotation significantly underestimate the true size of the BLR in the cloud model.}
    \label{fig:centroids}
\end{figure*}

The black hole mass and on-sky position angle are the physical parameters best constrained by our model, but one should note that the black hole mass is strongly correlated with inclination. If we restrict the sampler to inclinations of less than 45 degrees we find a best low inclination fit at $i =  37^{+4.6}_{-10}$ deg which then prefers a higher black hole mass of $M_\mathrm{BH} = 2.9^{+1.0}_{-1.3}\times 10^8 M_\odot$. In either case the inferred mean BLR size is of order $\sim 20 \mu$as. Taking the same distance and observed luminosity of 3C 273 as in G18 mean that the black hole mass as presented in Table \ref{table1} implies that the system is super-Eddington by a factor of a few \citep{luminosity}, but this can be rectified if we restrict the prior to low inclinations as shown above. Figure \ref{fig:centroids} shows how our model centroids compare to those in the data, in good agreement with ordered rotation around the jet. 

\newpage
\section{Discussion}

Radio observations of the jet indicate that the true inclination angle of 3C 273 is $\sim 20\pm{10}^\circ$\citep{jetAngle86}. \textbf{This disfavors the specific disk-wind model considered here, indicating the cloud model previously presented in G18 remains the best-fit model to the broad-line region in 3C 273. However we do not rule out disk-wind models completely, both for other AGN and/or in the case of different morphologies / kinematics such as those considered by \cite{ChajetHall2013, Flohic2012, Waters16, BaskinLaor2018, NaddafCzerny2022, Mathews2020} and others.} It remains to be explored whether other kinds of disk-wind models such as these can accurately fit GRAVITY data and reconcile the inclination discrepancy. 

If the disk-wind model presented here correctly described the BLR physics, our model would predict a black hole mass lower by a factor of $\sim 5$ compared with results previously published in G18, while implying an angular size of the broad-line region that is smaller by a factor of $\sim 2$. The reduced $\chi^2$ presented for the cloud model fit in G18 is given as $\sim 1.3$, which to two significant figures is slightly better than our reduced $\chi^2 \approx 1.35$. Still, both models appear to fit the data with roughly the same quality, and without external knowledge of the inclination angle this would imply possible systematic errors in estimating both the size of the BLR and the mass of the central black hole from interferometry data by factors of $\sim 2$ and $\sim 5$ respectively. 

\begin{figure}
    \begin{minipage}{0.47\textwidth}
    \includegraphics[width=1.\linewidth,keepaspectratio]{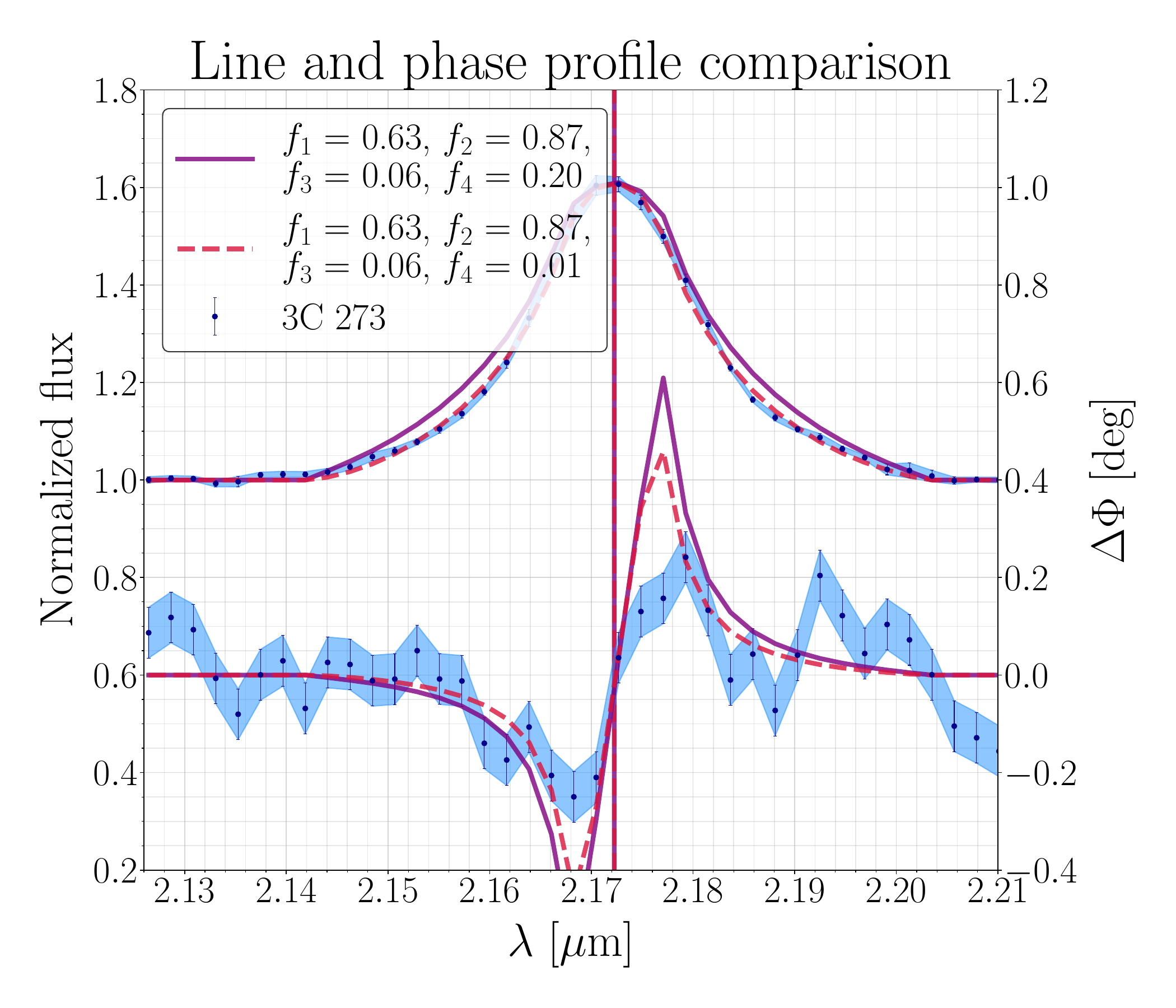}
    \caption{Similar to figure \ref{fig:best_fit}, only this time we show the average low inclination ($i\approx 30^\circ$) fit in dashed red, with the purple line showcasing the fine-tuning problem. To fit the data well the wind must be a Keplerian disk with a radial velocity gradient driven outflow only, including essentially only the $f_1$ and $f_2$ components of the model. Even small deviations from this finely tuned prescription degrades the fit significantly, as the solid purple line shows where $f_4$ is increased from $\approx 0$ to 0.2 and the fit quality clearly decreases.}
    \label{fig:finetune} 
    \end{minipage}
\end{figure}

If we use a prior that allows only for low inclination models we find a low inclination fit with a black hole mass that increases to $\sim 3\times 10^8 \ M_\odot$ in better agreement with the cloud model mass, but the on-sky size remains $\sim 20 \ \mu \textrm{as}$. This implies that the BH mass is roughly model independent and is instead simply strongly correlated with the inclination angle, a widely noted and expected correlation in astronomy. The BLR size is a model dependent systematic uncertainty, however, indicating that the models produce different values for the virial factor $f$ in equation \ref{RMeqn}. This low inclination best fit has a reduced $\chi^2 \approx 1.37$, which is slightly worse than the best fit at higher inclinations described but still acceptable. The average low inclination fit is much poorer, with a reduced $\chi^2 \approx 2.3$. This lower inclination best fit suffers from a problem of fine-tuning, however, as it essentially forces the $f_4$ and $f_3$ term to 0 in order to fit the data (in the low inclination case the average fit results for these terms are\textemdash with one $\sigma$ confidence intervals\textemdash $f_4 = 0.011^{+0.009}_{-0.01}$ and $f_3 = 0.06^{+0.013}_{-0.038}$ while $f_1 = 0.63^{+0.091}_{-0.22}$ and $f_2 = 0.87^{+0.048}_{-0.50}$). As figure \ref{fig:finetune} shows, the $f_1$ term must dominate the model in order to produce the broad single peak in the line profile\textemdash any deviation from a Keplerian thin-disk with a radial velocity outflow thus breaks the fit at lower inclinations. The low inclination fit to the data also fits the position angle signficantly worse, preferring $\theta_\mathrm{PA} = 257^{+1.9}_{-13}$$^\circ$ in disagreement with the measured value of roughly $220^\circ$. In comparing to G18, however, this result is interesting because the kinematics they consider in their cloud model allow for only Keplerian rotation / shears, thus the kinematics that are required by our low-fit model (radial and Keplerian shears only) are not entirely at odds with those required by the cloud model fit.

\begin{figure}
    \begin{minipage}{0.475\textwidth}
    \includegraphics[width=1.\linewidth,keepaspectratio]{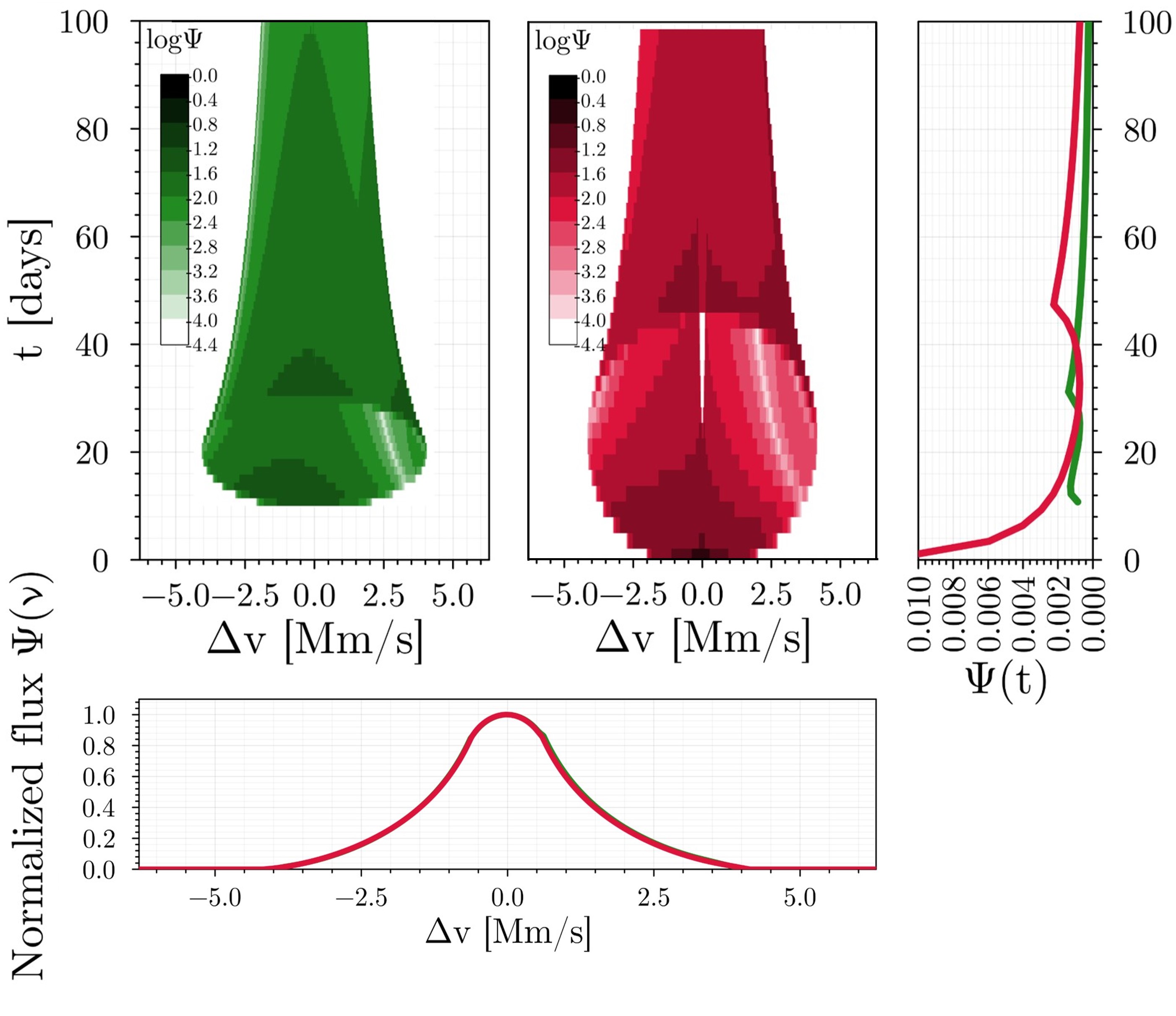}
    \caption{An echo image comparing our best fits at low and high inclinations (shown in figures \ref{fig:best_fit} and \ref{fig:finetune} respectively), which show the distribution of model intensity across resonant Sobolev surfaces in both frequency and time delay space. The low inclination model ($i\approx 30^\circ$) is shown at left in green and the high inclination model ($i\approx 90^\circ$) at right in red. In comparing to the figures in the appendix of Waters16 or figures 2, 4, and 5 in CM96 we see many similarities, but there are minor differences that come from our work as we consider additional terms in our fit (for example, the region on the right at low $t$ of our plot). Integrating the echo image along the time axis produces the line profiles shown in the bottom panel, while integrating across frequencies produces the transfer functions shown in the right panel.}
    \label{fig:RM} 
    \end{minipage}
\end{figure}

We also calculate an echo image of our best fit (shown in figure \ref{fig:RM}), to compare to reverberation mapping techniques and illustrate a mapping of the resonance conditions within our disk that produce the line and response profiles. Here we plot the change in frequency $\Delta v$ in units of Mm/s, i.e. $\Delta v = c\Delta\nu = -v_\phi\sin\phi\sin i$. The mean light travel time to our best fit value for $\bar{r}_{BLR}$ is $\lesssim 50$ days, in both the low or high inclination fit cases. 
The most recent results published by \cite{RM3c27319} have measured time delays of $\sim 145\pm{10}$ days in the rest frame of 3C273 using the $H\beta$ and $H\gamma$, while older results favor longer time delays of $\sim 300$ days \citep{GRAVITY7RM,GRAVITY8RM}. Previously published results in G18 produced a characteristic time delay of $\sim 145\pm{35}$ days at their value of $\bar{r}$ (corresponding to $\sim 50 \mu\textrm{as}$ on the sky), a size roughly twice as large as ours. While the time delay to our value of $\bar{r}$ is lower than this, this does not necessarily indicate that reverberation mapping data are in conflict with this model, as this value is weighted by the intensity and not responsivity of the disk. To do this would require a complete photoionization model of the BLR, and is complicated by many factors as described in \cite{Goad1993, GoadKorista2014} and papers cited therein. A better estimate could be made by convolving our model transfer function $\Psi(t)$ (as shown at right in Figure \ref{fig:RM}) with the continuum lightcurve of 3C 273, with the peak of the resulting cross-correlation function indicating the characteristic delay, but we leave this for future work. We include the delay number discussed here as a comparison to the previously published work in G18, and note that is roughly consistent to reverberation mapping data for such a rough estimate. In future work it may be possible to jointly fit reverberation mapping and GRAVITY data, in which case one would essentially fit for the transfer function in addition to the line profile and phase data.

Our model is simple, including only four possible wind launching terms simplified under the Sobolev approximation to rely on strong velocity gradients motivated by the local escape speed in the optically thick regime. It may be possible that further extensions/modifications of the model we use could help to mitigate the problems discussed here, but we leave this for future work. As it stands it appears that the simple disk-wind model considered in this work is in tension with the assumption that type 1 AGN are generally viewed at low inclinations, as we only expect to see single-peaked lines at higher inclinations (or the disk-wind model must be exceedingly well tuned at lower inclinations). Thus we plan to test different kinds of disk-wind models to see if this tension can be resolved, additionally including further resolved GRAVITY sources as well as reverberation mapping data in our fits.

\section{Conclusions}
We have fit a simple disk-wind launching model to GRAVITY data, showing that such a model can fit the line profiles, phase profiles, and emission centroids observed by GRAVITY. Our fit results in a smaller black hole mass for 3C 273 ($\approx 8\times 10^7 \ M_\odot$) than other modelling results, which would make the system likely super Eddington given its observed luminosity \citep{luminosity}, but this smaller inferred mass is a result of our fit preferring higher inclinations, which observations of the radio jet do not support. Our fit also prefers a smaller size for $R_{\mathrm{BLR}}$, which is independent of inclination and thus a model-dependent uncertainty. Combining these two uncertainties would lead to larger systematic errors in inferring the black hole mass (see equation \ref{RMeqn}), but we emphasize that (at least in the case of 3C 273) this disk-wind model is disfavored. 

In order to create a single peak in the line profile our fit requires the $f_1$ term be dominant over $f_3$ and $f_4$, which can be obtained by either forcing the viewing inclination of the system to be high or by fine-tuning the wind at low inclinations, such that only a Keplerian thin-disk with radial outflows is present. Overall our fit prefers a higher inclination, and our best fit value is much higher than what is inferred for 3C 273 from observations of the jet. Our fitting results prefer an inclination angle of $\sim 75^\circ$, while the jet for 3C 273 indicates an inclination angle of $\sim 20\pm{10}^\circ$ \citep{jetAngle86}. \textbf{Assuming the jet and the disk are not significantly misaligned, the cloud model presented by G18 appears to thus better match the data.} The phase and line profile fits in that work are at least as good as in our model but the cloud model produces a fit at much lower inclinations ($\sim 12^\circ$) as expected given the observational constraints on the jet orientation of 3C 273. We show that if we restrict the sampler to lower inclinations we obtain a fit that is plausible albeit slightly statistically worse than the higher inclination fit, but that this lower inclination fit suffers from a fine-tuning problem in the wind terms. 

Our model is simple and there may be extensions that further improve the fit we have not considered in this work, such as those considered by \cite{ChajetHall2013, Flohic2012, Waters16, BaskinLaor2018, NaddafCzerny2022, Mathews2020}. Furthermore it may be possible that not all quasar BLRs are governed by the same physics, and while 3C 273 may not be governed by disk-wind launching dynamics other quasars may still be. We hope to extend this work to include other disk-wind morphologies and kinematics, as well as more robustly test these models with reverberation mapping data in addition to GRAVITY data. Based on these initial results, however, it seems difficult to fit a disk-wind model such as this to any type 1 AGN as they are preferentially viewed at low inclinations. If the evidence continues to favor the cloud model as this result does, \textbf{it is of increasing importance to try to better understand the physical processes that can result in the BLR being best modelled as a distribution of cold and dense puffed up clouds of atomic gas, as well as how this picture can be connected to the strong observational evidence for outflows.}

\section{acknowledgments}
This work was supported in part by NSF grant AST-1909711 and an Alfred P. Sloan Research Fellowship (JD). We are grateful to the computing resources made available to us by Research Computing at CU Boulder, as this work utilized the Summit supercomputer, which is supported by the National Science Foundation (awards ACI-1532235 and ACI-1532236), the University of Colorado Boulder, and Colorado State University. The Summit supercomputer is a joint effort of the University of Colorado Boulder and Colorado State University. KL is especially grateful to Sajal Gupta for many helpful discussions over the course of the project, and to Marcel Corchado-Abelo for discussions on the rate of strain tensor. We are grateful to the anonymous referee for their thorough report, which greatly improved the quality of the paper. The code used in this work is available free and open-source on \href{https://github.com/kirklong/3C273DiskWindPaper}{GitHub}, and a plain-language summary of this work is available on the primary author's \href{https://www.kirklong.space/DiskWind3C273}{website}.

\facilities{VLTI(GRAVITY)\citep{GRAVITY17}}

\software{Julia,
          ptemcee,
          python
          }

\newpage

\bibliography{citations}{}
\bibliographystyle{aasjournal}

\appendix\vspace{-7mm}
\section{Full derivation of the line of sight velocity gradient}
As discussed in the text, the line of sight velocity gradient $\frac{\mathrm{d}v_l}{\mathrm{d}l}$ can be approximated with the rate of strain tensor. In computing $\frac{\mathrm{d}v_l}{\mathrm{d}l}\approx \hat{n}\cdot\boldsymbol{\Lambda}\cdot\hat{n}$ as given in \ref{eq1} we need the components of the rate of strain tensor $\Lambda_{ij}$ in spherical geometry, which we obtain from \cite{Batchelor68}. As in the text (see equation \ref{nhat}) we use: 
\begin{equation}\label{nhat2}
    \hat{n} = (\sin\theta\cos\phi\sin i + \cos\theta\cos i)\hat{r} + (\cos\theta\cos\phi\sin i - \sin\theta\cos i)\hat{\theta} - (\sin\phi\sin i)\hat{\phi}
\end{equation}

Using this we evaluate $\hat{n}\cdot\boldsymbol{\Lambda}\cdot\hat{n}$ as:
\begin{equation}\label{nhatLambdanhat}
\begin{split}
    \hat{n}\cdot\boldsymbol{\Lambda}\cdot\hat{n} = & \Lambda_{rr}\left(\sin\theta\cos\phi\sin i + \cos\theta\cos i\right)^2 +\Lambda_{\theta\theta}\left(\cos\theta\cos\phi\sin i - \sin\theta\cos i \right)^2+\Lambda_{\phi\phi}\left(\sin\phi\sin i\right)^2\\
    & +2\Lambda_{r\theta}\left(\sin\theta\cos\phi\sin i + \cos\theta\cos i \right)\left(\cos\theta\cos\phi\sin i - \sin\theta\cos i\right)\\
    & -2\Lambda_{\theta\phi}\left(\cos\theta\cos\phi\sin i -\sin\theta\cos i\right)\left(\sin\theta\sin i\right)-2\Lambda_{r\phi}\left(\sin\theta\cos\phi\sin i + \cos\theta\cos i\right)\left(\sin\phi\sin i\right)
\end{split}
\end{equation}

Applying the approximation that the disk is very thin and at the midplane we set $\theta=\frac{\pi}{2}$, reducing equation \ref{nhatLambdanhat} to: 
\begin{equation}\label{nhatLambdanhat2}
\begin{split}
    \hat{n}\cdot\boldsymbol{\Lambda}\cdot\hat{n} = & \Lambda_{rr}\left(\cos\phi\sin i\right)^2 -\Lambda_{\theta\theta}\left(\cos i \right)^2+\Lambda_{\phi\phi}\left(\sin\phi\sin i\right)^2\\
    & -2\Lambda_{r\theta}\left(\cos\phi\sin i\right)\left(\cos i\right)\\
    & +2\Lambda_{\theta\phi}\left(\cos i\right)\left(\sin i\right)-2\Lambda_{r\phi}\left(\cos\phi\sin i \right)\left(\sin\phi\sin i\right)\\
\end{split}
\end{equation}

Simplifying, this leaves us with: 
\begin{equation}\label{nhatLambdafinal}
    \hat{n}\cdot\boldsymbol{\Lambda}\cdot\hat{n} = \sin^2i\left(\Lambda_{rr}\cos^2\phi + \Lambda_{\phi\phi}\sin^2\phi-2\Lambda_{r\phi}\sin\phi\cos\phi\right) - \sin i \cos i \left(2\Lambda_{r\theta}\cos\phi - 2\Lambda_{\theta\phi}\sin\phi\right) + \Lambda_{\theta\theta}\cos^2 i
\end{equation}

From \cite{Batchelor68} the $\Lambda_{ij}$ terms in spherical coordinates are:
\begin{equation}\label{Lambdaterms}
\begin{gathered}
    \Lambda_{rr} = \frac{\partial v_r}{\partial r}\textrm{;  }\Lambda_{\theta\theta}=\frac{1}{r}\frac{\partial v_\theta}{\partial\theta} + \frac{v_r}{r}\textrm{;  }\Lambda_{\phi\phi} = \frac{1}{r\sin\theta}\frac{\partial v_\phi}{\partial \phi} + \frac{v_r}{r} + \frac{v_\theta\cot\theta}{r}\textrm{;  }\\
    \Lambda_{r\theta} = \frac{1}{2}\left(r\frac{\partial}{\partial r}\left(\frac{v_\theta}{r}\right)+\frac{1}{r}\frac{\partial v_r}{\partial \theta}\right)\textrm{;  }
    \Lambda_{\theta\phi} = \frac{1}{2}\left(\frac{\sin\theta}{r}\frac{\partial}{\partial \theta}\left(\frac{v_\phi}{\sin\theta}\right)+\frac{1}{r\sin\theta}\frac{\partial v_\theta}{\partial \phi}\right)\textrm{;  }
    \Lambda_{r\phi} = \frac{1}{2}\left(\frac{1}{r\sin\theta}\frac{\partial v_r}{\partial \phi} + r\frac{\partial}{\partial r}\left(\frac{v_\phi}{r}\right)\right)    
\end{gathered}
\end{equation}
Noting again that $\theta = \frac{\pi}{2}$, and assuming that the disk is axisymmetric in $\phi$ (thus all of the $\frac{\partial}{\partial\phi}$ terms are 0) gives:
\begin{equation}\label{Lambdaterms2}
\begin{gathered}
    \Lambda_{rr} = \frac{\partial v_r}{\partial r}\textrm{;  }\Lambda_{\theta\theta}=\frac{1}{r}\frac{\partial v_\theta}{\partial\theta} + \frac{v_r}{r}\textrm{;  }\Lambda_{\phi\phi} = \frac{v_r}{r} \textrm{;  }\\
    \Lambda_{r\theta} = \frac{1}{2}\left(r\frac{\partial}{\partial r}\left(\frac{v_\theta}{r}\right)+\frac{1}{r}\frac{\partial v_r}{\partial \theta}\right)\textrm{;  }
    \Lambda_{\theta\phi} = \frac{1}{2}\left(\frac{\sin\theta}{r}\frac{\partial}{\partial \theta}\left(\frac{v_\phi}{\sin\theta}\right)\right)\textrm{;  }
    \Lambda_{r\phi} = \frac{1}{2}\left(r\frac{\partial}{\partial r}\left(\frac{v_\phi}{r}\right)\right)    
\end{gathered}
\end{equation}
We now apply a final approximation to the disk, in which we assume that $v_r\approx v_\theta\approx0$ in keeping with the standard assumptions for thin disks. Plugging this result into equation \ref{nhatLambdafinal} then gives us:
\begin{equation}\label{nHatLambdaij}
\begin{split}
    \hat{n}\cdot\boldsymbol{\Lambda}\cdot\hat{n} =& \sin^2i\left(\frac{\partial v_r}{\partial r}\cos^2\phi -\left(r\frac{\partial}{\partial r}\left(\frac{v_\phi}{r}\right)\right)\sin\phi\cos\phi\right)\\
    &- \sin i \cos i \left(\left(r\frac{\partial}{\partial r}\left(\frac{v_\phi}{r}\right)+\frac{1}{r}\frac{\partial v_r}{\partial \theta}\right)\cos\phi - \left(\frac{\sin\theta}{r}\frac{\partial}{\partial \theta}\left(\frac{v_\phi}{\sin\theta}\right)\right)\sin\phi\right)\\
    &+ \frac{1}{r}\frac{\partial v_\theta}{\partial\theta} \cos^2 i
\end{split}
\end{equation}
We now turn our attention to the derivatives. We assume Keplerian orbits such that $v_\phi = \sqrt{\frac{GM}{r}}$ and thus $\frac{\partial v_\phi}{\partial r} =  \frac{-v_\phi}{2r}$. Applying this in conjunction with the chain rule, we can write: $r\frac{\partial}{\partial r}\left(\frac{v_\phi}{r}\right) = \frac{\partial v_\phi}{\partial r} - \frac{v_\phi}{r} = -\frac{3v_\phi}{2r}$. Similarly $r\frac{\partial}{\partial r}\left(\frac{v_\theta}{r}\right) = \frac{\partial v_\theta}{\partial r} - \frac{v_\theta}{r}$ and $\frac{\sin\theta}{r}\frac{\partial}{\partial\theta}\left(\frac{v_\phi}{\sin\theta}\right) = \frac{1}{r}\frac{\partial v_\phi}{\partial \theta} - \frac{v_\phi}{r}\cot\theta =  \frac{1}{r}\frac{\partial v_\phi}{\partial \theta}$. 

This reduces equation \ref{nHatLambdaij} to:
\begin{equation}\label{nHatLambdaij2}
\begin{split}
    \hat{n}\cdot\boldsymbol{\Lambda}\cdot\hat{n} =& \sin^2i\left(\frac{\partial v_r}{\partial r}\cos^2\phi + \frac{3v_\phi}{2r}\sin\phi\cos\phi\right)\\
    &- \sin i \cos i \left(\left(\frac{\partial v_\theta}{\partial r} - \frac{v_\theta}{r}+\frac{1}{r}\frac{\partial v_r}{\partial \theta}\right)\cos\phi - \left(\frac{1}{r}\frac{\partial v_\phi}{\partial \theta}\right)\sin\phi\right)\\
    &+ \frac{1}{r}\frac{\partial v_\theta}{\partial\theta} \cos^2 i
\end{split}
\end{equation}

We now must apply some kind of physical prescriptions to the remaining derivatives, which we do in keeping with the Sobolev approximation that there must be \textit{large} velocity gradients present. Using this we approximate the gradients in terms of local escape velocities, i.e. $\frac{\partial v_r}{\partial r} \approx 3\sqrt{2}\frac{v_\phi}{r}$, $\frac{\partial v_\theta}{\partial \theta} \approx \frac{v_{esc}}{\left(H/R\right)}$ (with $H/R\ll 1$ being the scale height of the disk), and $\frac{\partial v_\theta}{\partial r} \approx \frac{\partial v_r}{\partial r}$. Here we assume the local escape velocity is $v_e = \sqrt{\frac{2GM}{r}} = \sqrt{2}v_\phi$\textemdash this would seem to imply that $\frac{\partial v_r}{\partial r} = \sqrt{2}\frac{v_\phi}{r}$, but CM96 adopted an arbitrary extra factor of 3 (seemingly to assume the wind launching regions generate an outflow that travels at a substantial velocity with respect to the source) that we keep to better compare with their results. Since $v_\phi$ is  function of $r$ alone $ \frac{\partial v_\phi}{\partial \theta} = 0$, and finally we also set $\frac{\partial v_r}{\partial \theta} = 0$ in keeping with the idea of a geometrically thin disk. This allows us to arrive at the form presented in equation \ref{eq3} in the text:

\begin{equation} \label{eq3Appendix}
\boxed{\hat{n}\cdot\boldsymbol{\Lambda}\cdot\hat{n} \approx \frac{\mathrm{d}v_l}{\mathrm{d}l} = 3\frac{v_\phi}{r} \sin^2i\cos\phi\left[\sqrt{2}\cos\phi + \frac{\sin\phi}{2}\right]-\sin i \cos i \left[3\sqrt{2}\frac{v_\phi}{r}\cos\phi\right] +\cos^2i\left[\frac{1}{r}\frac{v_{esc}}{(H/R)}\right]}
\end{equation}

\textbf{Note that there is a sign difference between the term proportional to $\sin i\cos i$ than in similar work done by \cite{Flohic2012}, which is the result of different assumptions for the underlying kinematics / velocity gradient fields, but as noted therein this has minimal effect on the shape of the line profiles.}

\section{Full phase data}\vspace{-5mm}
\begin{figure}[ht!]
    \centering
    \includegraphics[width=0.9\textwidth,keepaspectratio]{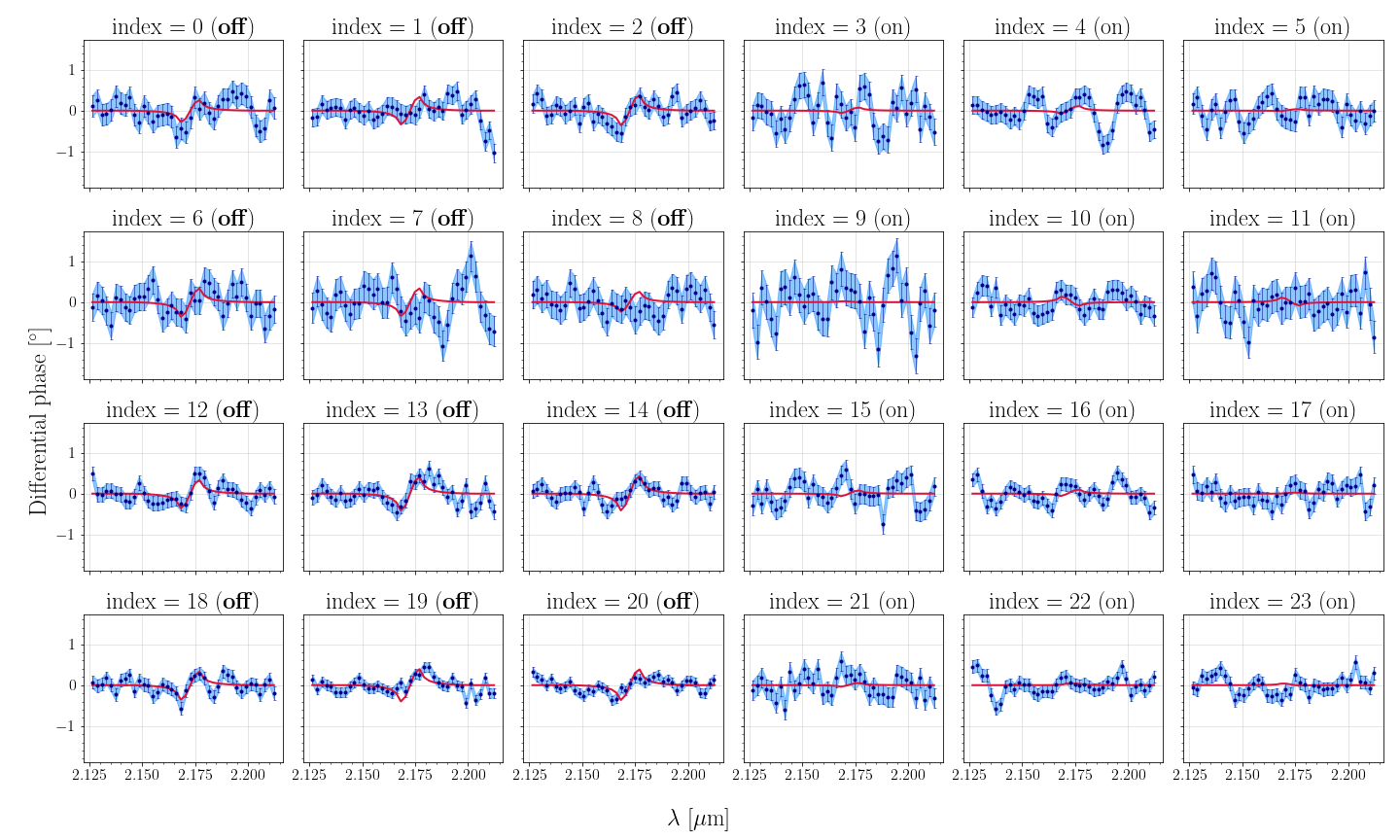}
    \caption{Individual phase profiles for all 24 possible configurations (6 baselines at 4 epochs). The ``off" axis (from the 3C 273 jet orientation) baselines are the ones that are averaged to create the figures shown in the text. Figure \ref{fig:baselines} below shows this alignment and why this choice is made. The red lines are the model phases from the average parameters given in table \ref{table1}.}
    \label{fig:fullphase}
\end{figure}

\begin{figure}[ht!]
    \centering
    \includegraphics[width=\textwidth,height=\textheight,keepaspectratio]{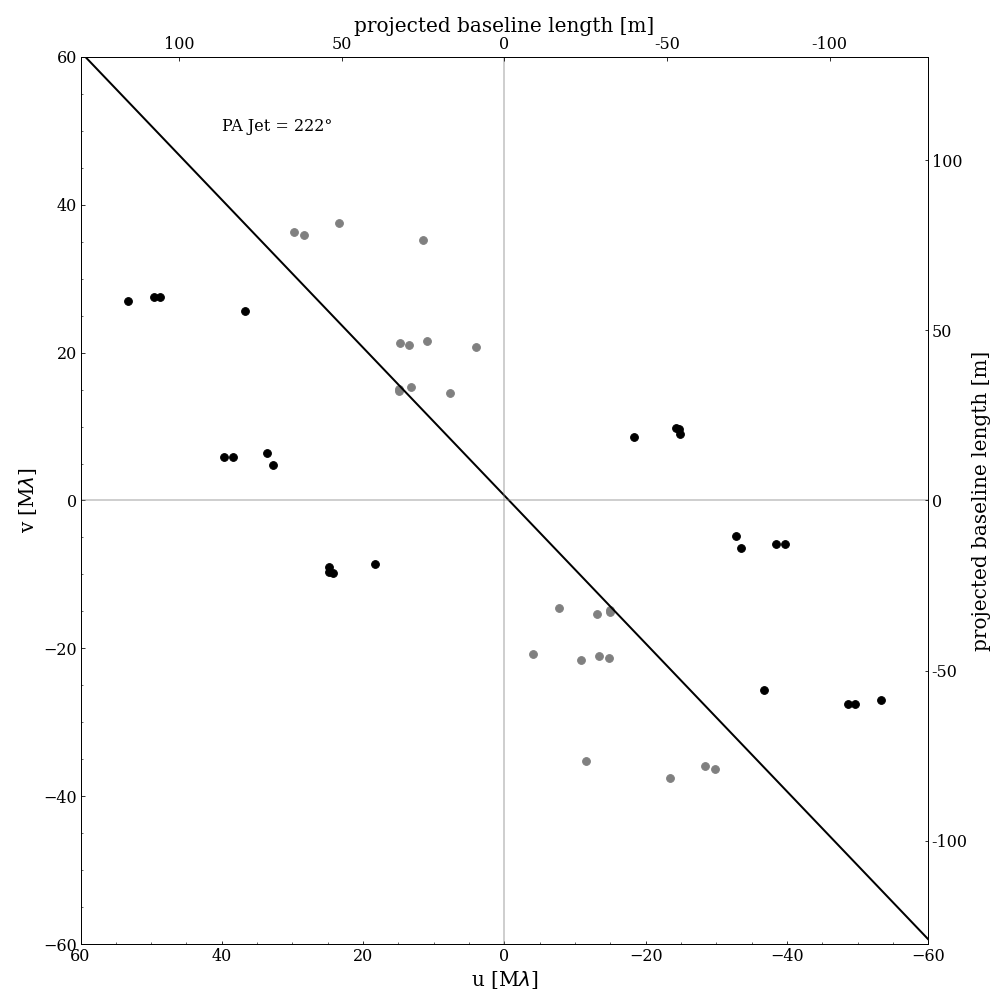}
    \caption{Here we show the baseline and epoch configurations, similar to E1 in G18. The black dots correspond to the bold ``off" axis phase plots in figure \ref{fig:fullphase} above, while the lighter grey markers correspond to the ``on" axis baselines. We only expect to detect significant asymmetries in the space off of the jet axis, and indeed we observe this, so we only include these off axis baselines in the plots shown in the paper. These on axis baselines are still included in the model, however, so all baselines are fit equally. As shown in figure \ref{fig:fullphase} the fit converges to an essentially flat line in phase space for the on axis baselines as expected for no ordered rotation signature in the jet itself.}
    \label{fig:baselines}
\end{figure}
\vspace{5mm}
\section{MCMC distributions}
\begin{figure}[ht!]
    \centering
    \includegraphics[width=\textwidth,height=\textheight,keepaspectratio]{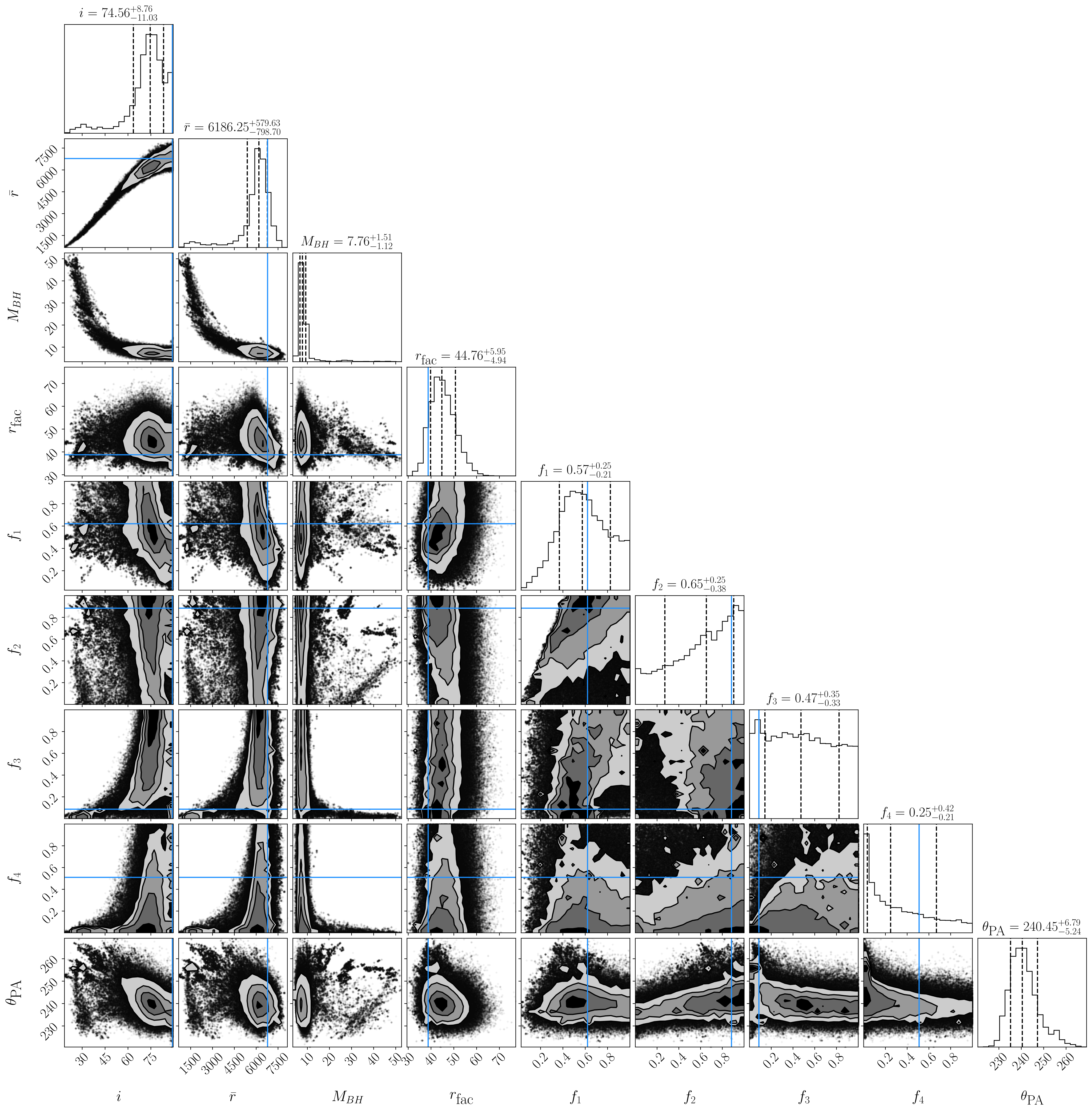}
    \caption{A corner plot showing the regions of parameter space explored and their dependences on one another, with 1D histograms for each. The units are the same as in Table \ref{table1}. Note that in the interest of space/readability the normalization parameter $n$ and the wavelength shift parameter $\Delta\lambda_c$ have been omitted as these are well constrained and the least physical of our parameters\textemdash a version of this plot with all the parameters is available online at the \href{https://github.com/kirklong/3C273DiskWindPaper/blob/main/fullCorner.png}{GitHub repository for this project}. The non-Gaussian shape of several of the histograms illustrates the importance of using multiple temperatures in the MCMC fitting to ensure the sampler does not get stuck in a local minimum. The blue lines indicate the best fit solution, which is essentially the high inclination model originally considered by Chiang and Murray in CM96, but note the difference in the reduced $\chi^2$ between the best fit and the average parameters is only $\sim 0.01$. While the various wind launching terms are poorly constrained, it is interesting to note that it appears (from the 1D histograms) that the sampler prefers $f_2$ approach 1 and $f_4$ 0. $f_2$ should be 1 in an ideal thin disk, as the Keplerian shear is non-negotiable. $f_4$ represents a form of isotropic emission which creates a double-horned profile, which means the contributions from it need to be small in order to preserve the observed single peak. The plot was created using \cite{corner}.}
    \label{fig:corner}
\end{figure}

\end{document}